\documentstyle[fake]{article}
\newbox\grsign \setbox\grsign=\hbox{$>$} \newdimen\grdimen \grdimen=\ht\grsign
\newbox\laxbox \newbox\gaxbox
\setbox\gaxbox=\hbox{\raise.5ex\hbox{$>$}\llap
     {\lower.5ex\hbox{$\sim$}}}\ht1=\grdimen\dp1=0pt
\setbox\laxbox=\hbox{\raise.5ex\hbox{$<$}\llap
     {\lower.5ex\hbox{$\sim$}}}\ht2=\grdimen\dp2=0pt
\def\gax{\mathrel{\copy\gaxbox}}

\begin{document}

\title{POST-T TAURI STARS: A FALSE PROBLEM}

\author{Francesco Palla\altaffilmark{1,2}
and Daniele Galli\altaffilmark{1}} 

\altaffiltext{1}{Osservatorio Astrofisico di Arcetri, Largo E.~Fermi 5,
 I--50125 Firenze, Italy}

\altaffiltext{2}{Laboratoire de Radioastronomie, Ecole Normale Superieure,
24 rue Lhomond, 75005 Paris, France}

\authoremail{palla@arcetri.astro.it, galli@arcetri.astro.it}

\begin{abstract}

  We consider the problem of the apparent lack of old T Tauri stars in
low-mass star forming regions in the framework of the standard model of low-mass 
star formation.  
  We argue that the similarity between molecular cloud lifetime and
ambipolar diffusion timescale implies that star formation does not take
place instantaneously, nor at a constant rate.
  We conclude that the probability of finding a large population of old
stars in a star forming region is {\it intrinsically} very small  and
that the post-T Tauri problem is by and large not existent.

\end{abstract}


\section{Introduction}

The problem of the possible incompleteness of the T Tauri population
was first addressed by Herbig (1978) who noted that there was an
embarrising deficit of T Tauri stars (TTS) of age greater than $\sim$ 2 Myr
in nearby star forming regions. Herbig introduced the possibility of a
missing population of ``post-T Tauri" stars with ages older than $\sim$
5$-$10 Myr and searches to identify such stars have been conducted
thereafter.  However, the census of the low-mass stellar population
is still a matter of debate.

Some authors argue that the combination of classical and weak TTS
observed {\it within} or {\it close to} molecular clouds represents the
whole stellar population in regions like the Taurus-Auriga complex
(e.g.  Hartmann et al. 1991; Gomez et al. 1992).  Other authors have
stressed the opposite viewpoint whereby the dominant stellar population
is still to be determined, since many young stars can be found at large
distances from the parent molecular clouds. In particular, Feigelson
(1996) has used the recent discovery made by the ROSAT All Sky Survey
of many weak TTS tens of parsecs away from nearby low-mass star forming
complexes to argue that the census is far from being complete because
of insufficient spatial coverage.  If the newly found stars are
genetically linked to those found within the boundaries of molecular
clouds, then the total number of low-mass stars formed in such regions
would increase dramatically. Assuming that the star formation has been
taking place at a constant rate over the last 10 million years (i.e.
during the typical lifetime of a molecular cloud), Feigelson (1996)
estimates that the predicted deficit of stars older than 2--5 Myr
exceeds hundreds or thousands of TTS. One obvious implication of this
conclusion is that the star formation efficiency of molecular clouds
would be much higher than the few percent currently estimated (e.g.
Cohen \& Kuhi 1979).

In this paper we take the opposite point of view and argue that the
lack of older stars in low-mass star forming regions is a consequence
of the star formation process itself rather than the dispersal of stars
outside well-surveyed regions. The main idea behind this statement is
that it is erronoeus to think of star formation as an instantaneous
process during the lifetime of a molecular cloud, but that rather it
takes a long time for establishing the proper initial conditions
conducive to gravitational collapse and star formation. We identify
this timescale with the ambipolar diffusion time in weakly ionized
media, as suggested by various authors (e.g. Mouschovias 1976; Shu,
Adams \& Lizano 1987, SAL).  It is also shown that this timescale is
similar to the estimated lifetime of molecular clouds and that
therefore the probability of finding a large population of older stars
is {\it intrinsically} very small. The star formation efficiency within
a star forming region is {\it not} constant as a function of time, but
it is very low initially and increses steeply at later times.
Nevertheless, the overall value of the efficiency, few percent, is
small since it is limited by the mechanisms that determine the survival
of a molecular cloud complex. We do not discard the possibility that a
certain fraction of young stars can indeed be found at large distances
from molecular clouds, but we dispute the conclusion that such a
population would outnumber the known TTS by the estimated factors
($10^2-10^3$).  Thus, we conclude that the post-T Tauri problem is
basically a false problem.

\section{Time scales for core formation and gravitational collapse}

In this section, we briefly review the main theoretical ideas behind
dense cloud core formation and ambipolar diffusion. Following other
authors, we believe that the formation of low-mass stars takes place
within cloud cores which are cold, $T\sim 10$~K, dense, $n_{H_2} \gax
10^4$~cm$^{-3}$, and weakly ionized, $x_{\rm i}\sim 10^{-8}-10^{-7}$
(e.g. Myers \& Benson 1989).  Measurements of the magnetic field
strength indicate the presence of fields of 10--50 $\mu$G (Myers \&
Goodman 1988), which could provide the bulk of the support to the cloud
cores. These clouds appear to be magnetically subcritical, i.e. their
mass-to-flux ratio is less than a critical value $(M_{\rm cl}/\Phi_{\rm
cl})_{\rm cr}\simeq 0.1/\sqrt{G}$, where $\Phi_{\rm cl}$ is the
magnetic flux threading the cloud, and $M_{\rm cl}$ its mass.

For a frozen-in magnetic field, subcritical clouds cannot undergo
gravitational collapse if the magnetic flux is conserved.  However,
collapse can occur by redistribution of magnetic flux by ambipolar
diffusion in a lightly ionized medium, where the ion-neutral
collisional coupling becomes less effective and magnetic forces drive
the ionized matter through the neutrals, thus decoupling the field from
the neutral matter (Mestel \& Spitzer 1956).  This process does not
alter the magnetic field threading a cloud but simply redistributes
mass among the central flux tubes of a cloud (Mouschovias 1976).  The
process of ambipolar diffusion is negligible in cloud envelopes because
of the relatively high degree of ionization ($x_{\rm i}\gax 10^{-5}$).

A good approximation for the density of ions, valid in a wide range of
molecular cloud densities, is $\rho_{\rm i}=C\rho_{\rm n}^{1/2}$, where
$\rho_{\rm i}$ and $\rho_{\rm n}$ are the ion and neutral densities,
respectively, and $C=3\times 10^{-16}$~cm$^{-3/2}$~g$^{1/2}$ is a
constant. Under these conditions ambipolar diffusion sets in gradually
and its characteristic timescale in a molecular cloud core of
size $R$ and density $\rho_{\rm n}$ is (e.g. Shu~1992)
$$
t_{\rm AD}\sim {R\over v_{\rm d}}\sim
{4\pi\gamma C\rho_{\rm n}^{3/2}R^2\over B^2} =13\;
\left({n_{H_2}\over 10^4\;{\rm cm}^{-3}}\right)^{3/2}
\left({R\over 0.2\;{\rm pc}}\right)^2
\left({B\over 30\;\mu{\rm G}}\right)^{-2}~{\rm Myr}, \eqno(1)
$$
where $v_{\rm d}$ is the ion-neutrals drift velocity and
$\gamma=3.5\times 10^{13}$~cm$^3$~g$^{-1}$~s$^{-1}$ is the ion-neutrals
drag coefficient.  For typical molecular cloud cores, the ambipolar
diffusion time is long compared to the dynamical free-fall time, by a
factor of order 8--17 (SAL, McKee et al.~1993). Indeed, Myers \&
Goodman (1988) have shown that, using a crude virial equilibrium model
of thermal and magnetic support against gravity for the core, the
timescale for ambipolar diffusion always exceeds 10 Myr for cloud radii
greater than $\sim 0.1$~pc and magnetic field strengths less than $\sim
50$~$\mu$G.

Fiedler \& Mouschovias (1993) have presented detailed calculations of
the formation and contraction of isothermal, self-gravitating, magnetic
cores in axisymmetric geometry over a wide range of densities. Their
models clearly show that, for the assumed reference physical quantities
($n_{\rm ref}\sim 10^3$~cm$^{-3}$, $B_{\rm ref}\sim 30~\mu$G), the
quasi-static contraction lasts about 15 Myr: at that point the central
mass-to-flux ratio has increased by a large factor (10--50) and exceeds
the critical value for gravitational collapse. 

In order to estimate the typical timescale for the subsequent phase of
gravitational collapse and protostellar accretion, we can use the basic
results obtained by Shu (1977) and Stahler, Shu \& Taam (1980) for the
case of the collapse of a singular isothermal sphere. In such scenario,
the mass accretion rate is given by $ \dot M_{\rm acc} = \alpha a_{\rm
eff}^3/ G $, where $a_{\rm eff}$ is the effective sound speed, and
$\alpha$ a constant of order unity. For a core at $T=10$~K that has
lost its magnetic support, the mass accretion rate is of order $\dot
M_{\rm acc}\sim$ few $\times 10^{-6}~M_\odot$~yr$^{-1}$.  Thus, the
typical accretion timescale (which is comparable to the free fall time)
is
$$
t_{\rm acc}\equiv {M_\ast\over \dot M_{\rm acc}}\sim 10^5-10^6~{\rm yr}
, \eqno(2)
$$
for solar mass stars. Such a time is much shorter than the core
formation timescale. Thus, in a given molecular cloud, by the time the
cores acquire the proper conditions to initiate gravitational collapse,
most of them must have already formed a low-mass star (e.g. Beichman et
al. 1986, Lizano \& Shu 1989).

The failure to recognize the existence of a long gestation period
preceding the actual stellar birth and infancy, is, in our opinion, the
basic misconception at the origin of the problem of post-T Tauri stars.

\section{Lifetime of molecular clouds}

In this section, we present estimates of the lifetimes of molecular
clouds. As it will be shown below, the main point is that such
lifetimes are of the same order of those for core formation that we
have just derived. Since the issue of cloud ages is not as rigourous as
that of cloud core formation, we will present only the main arguments
used to estimate ages of giant molecular clouds (GMCs) and dark cloud
complexes.

As for the GMCs, the current consensus is that they should not be much
older than the crossing times of their constituent clumps. Elmegreen
(1985) and Falgarone \& Puget (1986) have estimated that in the case of
clumpy magnetic molecular clouds, the lifetime is about 2--5 cloud
crossing times, or approximately $t_{\rm MC}\sim 30$~Myr using typical
cloud parameters ($M_{\rm cl}\sim 10^2~M_\odot$ and $R_{\rm cl}\sim
1$~pc).  These analyses are applicable to the initial phases of large
molecular complexes, prior to the formation of massive O and B stars.

A more direct estimate of cloud age can be obtained from the age of the
associated star clusters. Elmegreen (1985) shows that the total age of
a molecular cloud is within a factor of 2 or 3 of the age of the stars
that it contains. The ages of the OB associations are in the range
$t_{\rm MC}\sim 10$--20~Myr. For some systems, it is also possible to
follow the change in cloud properties as a function of the age of the
associated cluster/association. The large scale CO survey made by
Leisawitz, Bash \& Thaddeus (1989) around 34 open clusters has shown a
marked decline in the cloud mass within 25 pc, or 10 cluster angular
diameters, of clusters by an age of 5 Myr. Clusters older than $\sim$10
Myr do not have molecular clouds associated with them more massive than
a few times $10^3~M_\odot$, and by an age of $\sim 30$~Myr the clouds
have almost completely disappeared.  Now, since very few giant
molecular clouds do not contain a massive association, the first
massive O stars, which are responsible of the wreackage due to powerful
winds and radiative heating, must have been formed soon after the
formation of a cloud complex. Thus, the estimate of $t_{\rm MC}\sim$
10--20~Myr provides a measure of the lifetime of a typical molecular
complex.

Let us now turn to the case of dark cloud complexes such as
Taurus-Auriga, $\rho$ Ophiucus and Chamaeleon where the issue of cloud
dispersal is more relevant to the problem of the post-T Tauri stars. We
have already noted that empirically about half of the well studied
dense ammonia cores have collapsed to form low-mass stars. The age of
the deeply embedded objects is of the order of few 10$^5$~yr, but
Beichman et al. (1986) argue that they may be even younger, based on
the timescale for star-core separation and on the analysis histograms
of T Tauri ages. In addition, Myers et al. (1987) have estimated the
timescale for the dispersal of circumstellar obscuration by comparing
the frequency of occurrence of obscured and optically revealed TTS in
Taurus-Auriga. They derive obscuration times of order 10$^5$ yr,
comparable to the dynamical time of the associated CO outflows.
Finally, the very high incidence of the outflows among embedded sources
in cores, together with the large momentum and kinetic energy suggest
that cores can be disrupted or dispersed typically in a few $10^5$~yr
(Myers et al. 1988). In the absence of such energetic events and due to
the lack of nearby massive stars, the small clouds found in these
complexes can in principle last for a very long time.  However, there
is no direct evidence for such long lives at present, and even if the
above arguments have overestimated the actual efficiency of the core
dispersal mechanism(s), it is reasonable to assume that such entities
have lifetimes no longer than few Myr.

To summarise, we have shown that typical lifetimes for cloud complexes
vary between few and few tens of Myr, with the latter being a more
realistic estimate. Considering the results of the previous section,
we can define a total time for star formation $t_{\rm SF}=t_{\rm AD}+t_{\rm
acc}+t_{\rm disp}$, and since the last two terms are much smaller than
$t_{\rm AD}$, we derive the following inequality
of timescales:
$$
t_{\rm MC}\sim 10--20~{\rm Myr}\gax t_{\rm SF}\sim t_{\rm AD}\sim 10~{\rm Myr}~\gg
t_{\rm acc}\sim t_{\rm disp}\sim 1~{\rm Myr}, \eqno(3)
$$

\noindent
where $t_{\rm MC}$, $t_{\rm AD}$, $t_{\rm acc}$, and $t_{\rm disp}$
represent the cloud lifetime, the timescale for core formation by
ambipolar diffusion, the timescale for protostellar accretion, and the
timescale for core dispersal. We will now discuss the implications of
such a result.

\section{Implications}

In addressing the problem of the post- (or older) TTS, the usual
assumption is made that {\it if} star formation has been continuous at
the present rate for at least 10 Myr, then the samples of known TTS
represent only a small fraction of the actual population. Feigelson
(1996) predicts that hundreds or thousands of such stars should exist
and concludes by saying that ``the mistery of the missing older TTS is
more dramatic today that when Herbig (1978) first raised the issue". In
fact, the age distributions of TTS lacks such older objects while, if
anything, they indicate a tendency for stars to appear very young. In
modern terms, many stars in the H-R diagram appear at, or very close
to, the stellar birthline for low- and intermediate-mass stars (Stahler
1983; Palla \& Stahler 1990). Hartmann et al. (1991) have noticed that
in the case of the Taurus-Auriga complex ``star formation has been
synchronized to a remarkable degree". From the analysis of other star
forming regions, Feigelson (1996) concludes that this synchronism
should be extended to clouds in all directions around the Sun, a fact
that is considered implausible given the variety of internal and
external conditions in each of them.

Other possible explanations, involving short-lived star formation,
episodic star formation, uncertainties in age determinations due to
poorly known evolutionary tracks, survey incompleteness cannot account
for the sharp decline in population for ages greater than $\sim 5$~Myr.
The suggestion that older TTS exist in large numbers well outside the
currently known star forming regions, and the evidence for such a
population provided by the ROSAT All Sky Survey represent therefore an
almost natural explanation to the problem. 

However, the picture of star formation we have outlined in the previous
sections indicates that in fact there is no need to invoke such a large
population of older stars and that, pushed it to the extreme
consequences, a problem of the missing population does not exist at
all.

A potential objection to this picture concerns the observed paucity of
{\it inactive} (i.e. not forming stars) molecular clouds relative to
{\it active} clouds in the solar neighborood. If the duration of the
star formation phase is much shorter than cloud's typical lifetimes,
then the ratio $R$ of active/inactive molecular clouds may become very
small:  for example, Feigelson (1996) finds that multiple short bursts
of star formation would lead to $R\sim 1:5$ -- $1:25$, in contradiction
with the observational evidence that only few clouds in the solar
neighborhood are inactive. We do not address here the issue of the {\it
duration} of the phase of copious star formation in a molecular cloud;
rather, we want to stress the fact that the time {\it required} for the
onset of this phase may represent, under typical conditions, a
substantial fraction of the entire cloud's lifetime.  Let us define the
fraction $A$ of active molecular clouds as $A\equiv (t_{\rm MC}-t_{\rm
SF})/t_{\rm MC}$, and let $I\equiv t_{\rm SF} /t_{\rm MC}$ be the
fraction of inactive clouds, such that $A+I=1$.  Therefore $$ R\equiv
{A\over I} = {t_{\rm MC}-t_{\rm SF}\over t_{\rm SF} }, \eqno(4) $$ so
that if we accept the suggestion that $t_{\rm MC}\gax t_{\rm SF}$, then
the ratio $R$ of active to inactive clouds becomes of order unity, or
larger. With our fiducial values $t_{\rm MC}\simeq 20$~Myr, $t_{\rm
SF}\simeq 10$~Myr, a fairly comfortable ratio $R\simeq 1:1$ is
obtained.  Failing to grasp the essential fact that {\it star formation
is a slow process}, one would reach the opposite conclusion that $R\ll
1$.

There is another aspect of the observations that our model should also
address. In the same star forming regions, old stars with ages 5--30
Myr are indeed found, albeit not with the frequency predicted by the
constant star formation models. If star formation is a slow process,
how can we explain their presence?  Returning to the definition of
$t_{\rm AD}$, we see that this time scale depends rather sensitively on
the local physical conditions of the medium. In a given molecular
cloud, spatial fluctuations of e.g.  the magnetic field strength or
geometry, or ionization fraction, can induce naturally a spread in star
formation times. Considering all the rather sensitive dependence of
the coefficient $C$ in eq.~(1) on the metal depletion of the ambient
gas, that may well lead, in very dense and depleted regions, to a value
smaller by an order of magnitude than our adopted value (Umebayashi \&
Nakano~1990).  Thus, many different kinds of spatial fluctuations may
introduce a natural spread in the formation times of the cores and
hence in the age distribution of the stellar population. This spread is
well documented both in low-mass star forming regions and in young
stellar clusters.

A broader implication of the scenario we have outlined is that the star
formation rate of molecular clouds is not constant nor continuous in
time, but a steeply increasing function of time.  The rate in the
early phases of the life of a cloud complex is very low, because most
of the gas is at such low densities that gravitational collapse cannot
occur, and becomes quite large at later times (after the {\it typical}
ambipolar diffusion timescale).  However, the short timescales
associated with cloud destruction act as a self-regulatory mechanism to
keep the overall rate at the low observed values of few percent. Unlike
models based on the dispersal of TTS, we do not need to invoke a
revision of the low efficiencies originally estimated by Cohen \& Kuhi
(1979), nor a higher proportion of stars born in bound open clusters.
In other words, we agree with the viewpoint that the current census of
the young stellar population is close to being complete, and there is
no need to invoke a large population of missing young stars.

We are thus left with the last question: what are then the dispersed T
Tauri stars detected by ROSAT? We partake the view that {\it some}
stars can escape from their parent clouds both because of the velocity
dispersions they inherit from the gas and because of dynamical
ejections due to few body encounters during the formation stages. But
we disagree with the interpretation that these stars are the majority
of those detected by ROSAT.  In our view, it is more likely that these
stars represent a mixture of older stars resulting from a previous
episode of large scale star formation, and of younger stars formed tens
of parsecs away from major clouds within cloudlets that have been
dissipated. Indeed, some X-ray sources detected far away from the
$\rho$ Ophiucus core have been tentatively identified with young
Class~I stellar objects (Casanova et al.~1995), coeval with the
majority of the cluster stars. Future observations will unravel the
true origin and nature of this interesting class of stars.

\bigskip
We would like to acknowledge useful discussions with the participants
at the IV French-Italian meeting on Star Formation, organized by Dr.
Th. Montmerle, during which most of the ideas presented here have been
discussed. This project has been funded by CNR grant n.~96/00317.

\end{document}